\documentclass{article}

\usepackage{graphicx}

\usepackage[nonatbib, preprint]{neurips_2019}
\usepackage[utf8]{inputenc} 
\usepackage[T1]{fontenc}    
\usepackage{hyperref}       
\usepackage{url}            
\usepackage{booktabs}       
\usepackage{amsfonts}       
\usepackage{nicefrac}       
\usepackage{microtype}      
\usepackage{amsmath,amssymb,tabularx}
\usepackage{algorithm}
\usepackage[noend]{algpseudocode}
\usepackage{caption}
\usepackage{amsmath}
\usepackage{amsthm}
\usepackage{tikz}
\usepackage{bm,braket}
\usepackage[T1]{fontenc}
\usepackage{bm}
\usepackage{authblk}

\usepackage{url}
\DeclareMathOperator{\E}{\mathbb{E}}
\def\bibfont{\footnotesize}

\title{Differentiable Molecular Simulations for Control and Learning}

\author[1]{Wujie Wang}
\author[1, 2]{Simon Axelrod}
\author[1]{Rafael G{\'o}mez-Bombarelli}

\affil[1]{  Department of Material Science and Engineering\\

  Massachusetts Institute of Technology\\
  
  Cambridge, MA 02139, USA \authorcr
  \{\tt wwj, rafagb\}@mit.edu}
  
\affil[2]{  Department of Chemistry and Chemical Biology\\

  Harvard University\\
  
  Cambridge, MA 02138, USA  \authorcr
  \{\tt simonaxelrod\}@g.harvard.edu}

\begin{document}
\maketitle
\begin{abstract}
Molecular dynamics simulations use statistical mechanics at the atomistic scale to enable both the elucidation of fundamental mechanisms and the engineering of matter for desired tasks. The behavior of molecular systems at the microscale is typically simulated with differential equations parameterized by a Hamiltonian, or energy function.  The Hamiltonian describes the state of the system and its interactions with the environment.  In order to derive predictive microscopic models, one wishes to infer a molecular Hamiltonian that agrees with observed macroscopic quantities.  From the perspective of engineering, one wishes to control the Hamiltonian to achieve desired simulation outcomes and structures, as in self-assembly and optical control, to then realize systems with the desired Hamiltonian in the lab. In both cases, the goal is to modify the Hamiltonian such that emergent properties of the simulated system match a given target. We demonstrate how this can be achieved using differentiable simulations where bulk target observables and simulation outcomes can be analytically differentiated with respect to Hamiltonians, opening up new routes for parameterizing Hamiltonians to infer macroscopic models and develop control protocols. 
\end{abstract}

\section{Introduction}
At the atomic level, physical processes are governed by differential equations containing many degrees of freedom. Macroscopic phenomena in matter emerge from microscopic interactions that can be simulated through numerical integration of the equations of motion. In classical simulations, these equations of motion are derived from a Hamiltonian quantity $H$. In quantum simulations, they are derived from a Hamiltonian operator $\hat{H}$.
Examples of microscopic quantities from simulations are time series of positions, velocities, and forces on atoms and molecules. From these, a rich family of macroscopic observables can be calculated to describe the configurational and temporal correlation functions of atoms. These observables determine different properties of the simulated materials.

Classically, simulating the positions of particles with conserved energy requires integrating the Hamiltonian equations of motion: 

\begin{equation} \label{eq:EOM}
 \begin{gathered}
 \frac{dp_i}{dt} = - \frac{\partial H}{\partial q_i}, \; \; \; \; \; \;
 \frac{dq_i}{dt} = \frac{\partial H}{\partial p_i},
 \end{gathered}
\end{equation}

where $p_i$ and $q_i$ are the respective momentum and position of the $i^{\mathrm{th}}$ particle. $H$ is the Hamiltonian of the system. For conservative systems, it is given by the sum of kinetic energy and the potential energy,

\begin{equation} \label{eq:Hamiltonian}
    H(\mathbf{p}, \mathbf{q}) = U(\mathbf{q}) + \sum_i^N \frac{p_i^2}{2 m_i},
\end{equation}
where boldface denotes the set of quantities for all particles, $U(\mathbf{q})$ is the potential energy and $p_i^2/(2m_i)$ is the kinetic energy of the $i^{\mathrm{th}}$ particle. 

Simulating an entire system explicitly tracking  all relevant degrees of freedom is computationally intractable in most cases. Typically, one is interested in a small subset of a system, such as a molecule or protein, and concerned only with the influence of the environment, also known as \textit{bath}, on the system but not the details of the environment itself. For this reason, one usually incorporates an environment Hamiltonian $H_b$ with coarse-grained macroscopic variables of interest into the original Hamiltonian: $H_{tot} = H + H_b$. The inclusion of $H_b$ is important in simulating systems under certain thermodynamic conditions. For example, the crystallization and melting of water occur under specific temperature and pressure. These conditions are imposed by the environment, which must therefore be incorporated into $H_b$. The environment, and therefore $H_b$, can also be explicitly controlled and optimized in many complex physical processes. \cite{Rotskoff2018, Tafoya2019, Sivak2016} For example, $H_b$ can represent an external laser that is varied to control chemical reaction dynamics. In nucleation processes, time-dependent temperature protocols can be designed to drive molecular self-assembly \cite{Sajfutdinow2018, Jacobs2015a} to escape kinetic traps. 

The control setting in molecular simulations is similar to that of robotics control of fine-grained degrees of freedom, as described by Zhong \textit{et al}. \cite{Zhong2019} However, molecular simulation is typically concerned with coarse-grained variables that control macroscopic thermodynamics states. These variables represent ensembles of many micro-states. Such control is usually realized by incorporating a dissipative Hamiltonian term.

Recent advances in differential equation solvers have shown that \textit{differentiable simulations} may be performed, in which the result of a simulation can be analytically differentiated with respect to its inputs. \cite{Chen2018, Lu2019, Li2019, Liang2019cloth, Holl2020, Lu2019pde, Long2017} This work applies differentiable simulation to a variety of molecular learning and control problems. We show that a Hamiltonian can be learned such that the macroscopic observables computed from a simulation trajectory match a given target. This is done through automatic differentiation of the macroscopic observables with respect to the system Hamiltonian. Moreover, we show that the same principles can be used to control the system Hamiltonian to force the system towards a target state. The examples mentioned in the paper can be found at \url{https://github.com/wwang2/torchmd}. 

\begin{figure}[ht]
  \centering
  \includegraphics[width=4in]{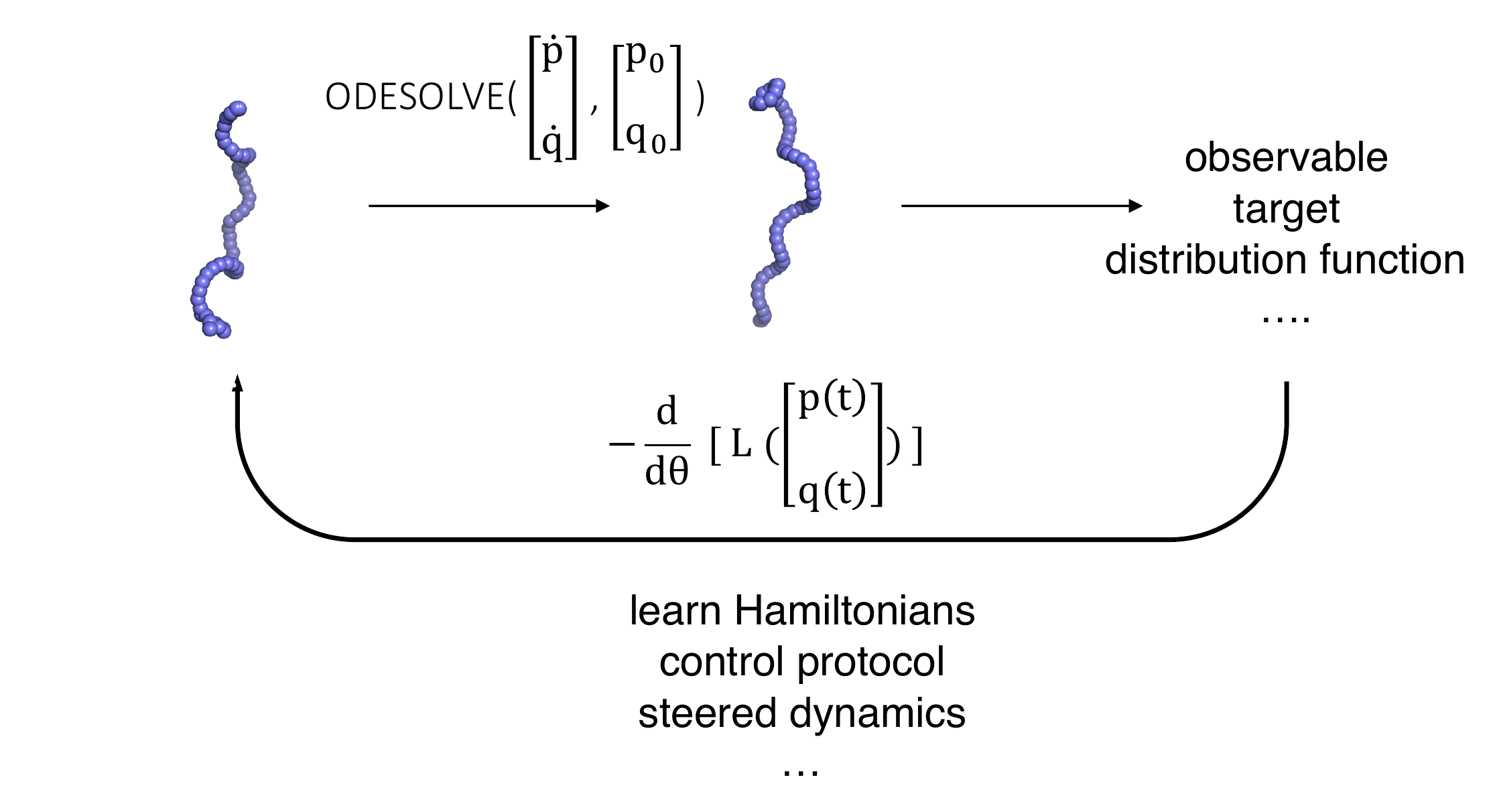}
  \caption{Differentiable molecular simulation workflow for learning and controlling physical models. The ODE is propagated with Eq. \ref{eq:EOM}; $p$ and $q$ denote the set of momenta and positions for all particles. }
  \label{fig:workflow}
\end{figure}

\section{Approach}
\subsection{Molecular simulations}

\textbf{Molecular simulations with a control Hamiltonian} In this work, we demonstrate the use of automatic differentiation in molecular dynamics simulations. To simulate systems under fixed thermodynamic conditions with macroscopic thermodynamic controls, most applications require fixed temperature, pressure, or strain. A typical way to impose thermodynamic constraints is to introduce modified equations of motion that contain virtual external variables. For example, to control the temperature of a systems, a virtual heat bath variable is needed to couple to the system variables. \cite{Nose1984, Martyna1992, Parrinello1982} These modified equations of motion can be integrated using ODE solvers. In the Appendix we provide details of the Nose-Hover chain, a common method for imposing constant temperature conditions. In our experiments we have implemented a differentiable version of this integrator to represent realistic dynamics of particle systems at constant temperature while allowing backpropagation.

The use of a graph neural network to represent a control Hamiltonian $H_b$ is also demonstrated in an example application. Graph neural networks (GNN) are a state-of-the-art architecture to learn the molecular potential energy function \cite{Gilmer2017, Schutt2017tensor, duvenaud_convolutional_2015} while preserving physical symmetries (translational, rotational and permutational). GNNs have shown flexible fitting power to represent molecular energies, and can be used to model control Hamiltonians (see Appendix). 

\textbf{Quantum dynamics with control}
To extend the notion of Hamiltonian dynamics to quantum mechanics, we consider the evolution of the wave function $\psi(\mathbf{x}, t) =e^{-i\int_{0}^{t} dt' \hat{H}(t')}\psi(\mathbf{x}, 0)$, where $\mathbf{x}$ are the spatial degrees of freedom, $\hat{H}$ is the Hamiltonian operator, and the exponential term is a unitary propagator. In abstract vector notation, this can be written as $\ket{\psi(t)} = e^{-i\int_{0}^{t} dt' \hat{H}(t')} \ket{\psi(0)}$. Time-dependent control terms in the Hamiltonian can be used to steer the evolution toward a desired outcome. For example, one can
manipulate the intensity spectrum and/or the phase of light to control quantum behaviour. The latter is known as coherent control, \cite{brumer_shapiro_book, rabitz_1992} and has been used to control the out-of-equilibrium behavior of various physical processes.  \cite{hi_coherent, qd_coherent, mol_coherent, gaas_coherent, rabitz_dream_alive} A prototypical light-controlled process is the  isomerization of the molecule retinal, which is responsible for human vision, \cite{vision_overview} light-sensing, \cite{phototaxis} and bacterial proton pumping. \cite{proton_pump} Changes in the light impingent on the molecule can alter the efficiency, or quantum yield, of isomerization. Coherent control of retinal within an entire protein, bacteriorhodopsin, was demonstrated in Ref. \cite{miller_coherent}. Further, coherent control of the isomerization of retinal was demonstrated \textit{in vivo} in the light-sensitive ion channel channelrhodopsin-2. \cite{in_vivo_control} This control led to the manipulation of current emanating from living brain cells. 

Computational and theoretical analysis of control can explain and motivate experimental control protocols. \cite{cyrille_inv_vivo, cyrille_coherent_bounds, cyrille_resonances, leonardo_mechanisms} In Ref. \cite{cyrille_resonances}, for example, an experimental control protocol was reproduced computationally and explained theoretically. A minimal model for retinal \cite{hahn_stock} was used to simulate the quantum yield (isomerization efficiency), and a genetic algorithm was used to shape the incident pulse.  With the minimal retinal model used commonly in the literature, \cite{simon_2, tscherbul_1, tscherbul_2, cyrille_resonances} we here analyze the control of the incident pulse through differentiable simulations. We allow for control of both the phase and intensity spectrum of the light; restriction to phase-only (\textit{i.e.} coherent) control is also straightforward. In particular, we show that back-propagation through a quantum simulation can be used to shape the incident pulse and control the isomerization quantum yield. This provides an example of out-of-equilibrium molecular control with differentiable molecular simulations. 

\subsection{Back-propagation with the adjoint method}

To be able to differentiate molecular simulations to reach a control target, we adopt the reverse-mode automatic differentiation method from the work by Chen \textit{et al.}, which uses adjoint sensitivity methods. \cite{Pontryagin1962, Chen2018}  Taking derivatives requires computation of the adjoint state $a(t) = dL/d(p(t),q(t))$. Evaluating the loss requires the reverse-time integration of the vector-Jacobian product:
\begin{equation}
    \frac{dL}{d\theta} = \int_{t_{i+1}}^{t_{i}} a(t) \frac{df(q(t), p(t), \theta) }{d\theta} dt,
\end{equation}
where $f(q, p, t)$ represents the Hamiltonian ODE defined in Eq. \ref{eq:EOM}, with $\theta$ being the learnable parameters in the Hamiltonian. The reverse-mode automatic differentiation computes the gradient through the adjoint states without backpropagating through the forward computations in the ODE solver. This has the advantage of having constant memory costs. The ability to output positions and momenta at individual timesteps allows one to directly compute observables and correlation functions from a trajectory. Separate reverse adjoint integrations are performed for observations at different individual observation times.

\section{Targeted molecular dynamics}

In this section, we show that with differentiable simulations, it is possible to bias the molecular dynamics in simulations towards a target state specified by a set of observables. From the perspective of engineering, a control Hamiltonian and thermodynamic protocol can be used to actively drive self-assembly of molecular structures. \cite{Nguyen2016, Jacobs2015, Jacobs2015a, Hormoz2011} This type of protocol connects a starting state to a final state, with the motion governed by two Hamiltonians. Being able to devise such simulation protocols is also useful in the context of non-equilibrium work relations (such as the celebrated Jarzynski equality \cite{Jarzynski1997a, Jarzynski1997} and Crook's Fluctuation Theorem, \cite{Crooks1999}), because this allows one to compute free energy differences with both equilibrium \cite{Vaikuntanathan2008} and non-equilibrium simulations. \cite{Rotskoff2018} A similar equality is also known in statistics as \textit{annealed importance sampling}. \cite{Neal1998, Grosse2015, Habeck2017} We show that one can adaptively learn a control Hamiltonian $H_b$ by perturbing a prior Hamiltonian with backpropagation to produce a state with a set of target observables and simulation outcomes.

\begin{figure}[ht]
  \centering
  \includegraphics[width=5in]{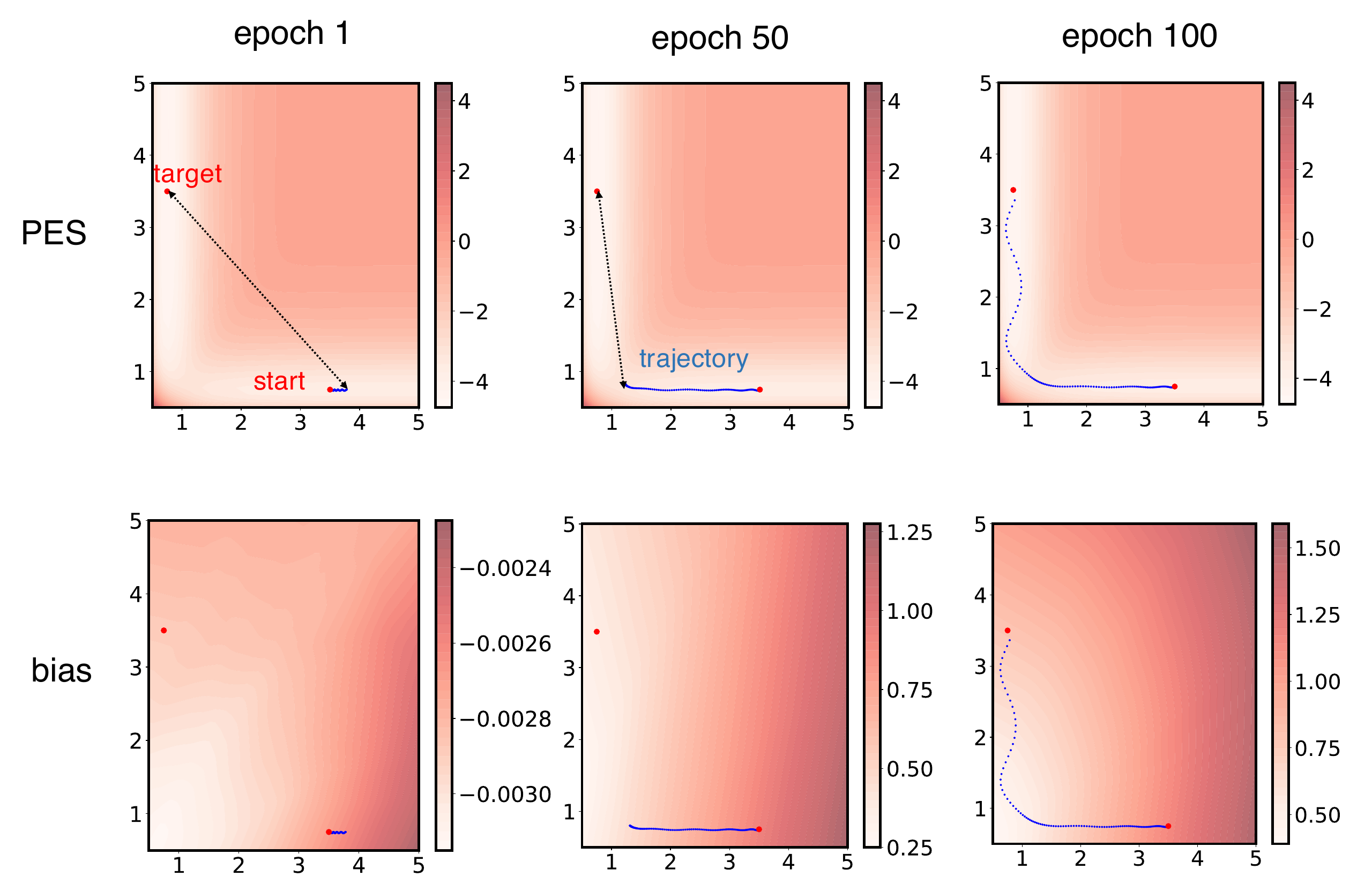}
  \caption{Controllable simulations in a toy 1 particle system in a 2D space. The control Hamiltonian $H_b$ is learned so that the particle goes from a starting position to a target. We use an MLP to parametrize a 2D bias potential. The target loss to minimize is the distance between the final and target positions. The particle is initialized with zero velocity. During training, we run Hamiltonian dynamics with 100 steps and compute the loss to update the bias potential and steer the particle trajectory toward the target state.}
  \label{fig:2d}
\end{figure}

\begin{figure}[ht]
  \centering
  \includegraphics[width=5in]{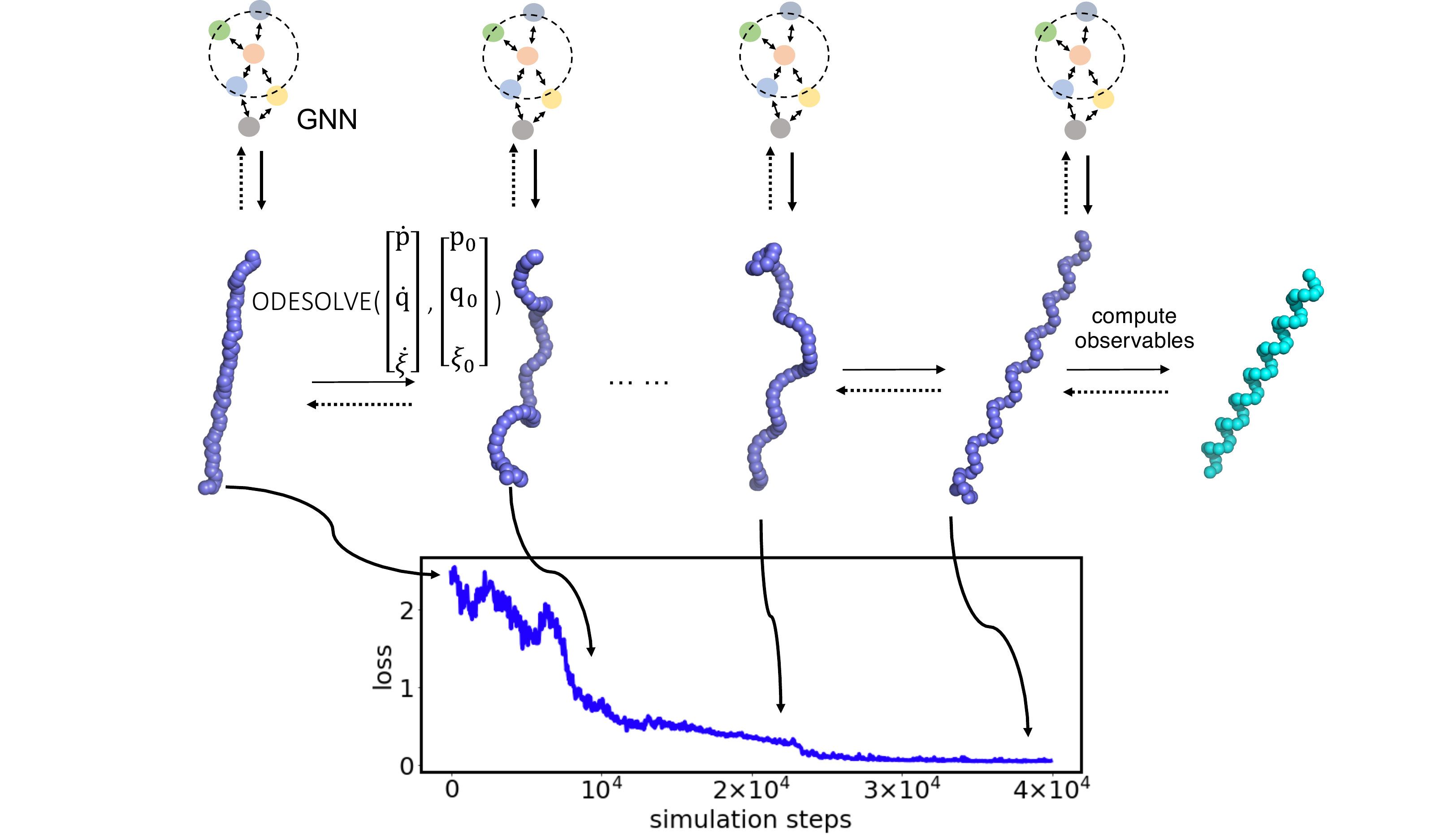}
  \caption{\textbf{A} We perform continuous model training during the simulations to bias a harmonic polymer chain toward a targeted helix shape. This is accomplished by training a bias Hamiltonian parameterized by a GNN. Here, $p$ and $q$ denote the set of momenta and positions for all particles, and $\xi$ denotes the Hamiltonian control variables that maintain a certain temperature. The simulations are run for 40000 steps, and the loss is computed and differentiated to update GNN weights every 40 simulation steps.}
  \label{fig:fold}
\end{figure}

In a first example, it is shown how differentiable molecular simulations allow control of dynamics by pertubatively seeking transitions pathways from a starting state to a target state. This learning scheme does not require a user-defined reaction coordinate. This is advantageous, because reaction coordinates are challenging to define in many complex dynamic processes, such as protein folding and chemical reactions. We demonstrate the application in toy 2-D and 3-D systems (Fig. \ref{fig:2d} and \ref{fig:fold}, respectively). In the 2D case, a bias potential is trained to steer a particle in 2D double well potential energy surface toward a target state. It starts as a small pertubative potential and then slowly increases in magnitude to help the particle pass over the barrier separating the states. This similar to biasing methods that have been used in identifying pathways in protein folding \cite{Mendels2018} and chemical reactions. \cite{Maeda2016}

To train a controllable 3D polymer folding process, we use a GNN to represent $H_b$, which biases a linear chain of identical particles connected with a harmonic Hamiltonian into a helical fold (see Fig. \ref{fig:fold}). The \textit{prior} potential is a sum over identical harmonic bond potentials, with an equilibrium bond length of 1 in arbitrary units: $U_{prior} = \sum_k (r_k - 1)^2 $, where $k$ is the index for the bonds along the chain. This Hamiltonian is perturbed in the simulations to produce a folded helical structure at a constant temperature using the Nose-Hoover Chain integrator described above and in the Appendix.  We back-propagate through the simulations to continuously update the GNN parameters, so that the loss function $L = \sum_i^{observables} ( \phi_i(q(t_1)) - \phi_i(q_{helix}) )^2$ is minimized. Here, the set of functions ${\phi_i}$ include structural variables of the polymer chain: bond distances, angles and dihedrals angles along the chain.

\section{Learning from observables}

\begin{figure}[ht]
  \centering
  \includegraphics[width=5.5in]{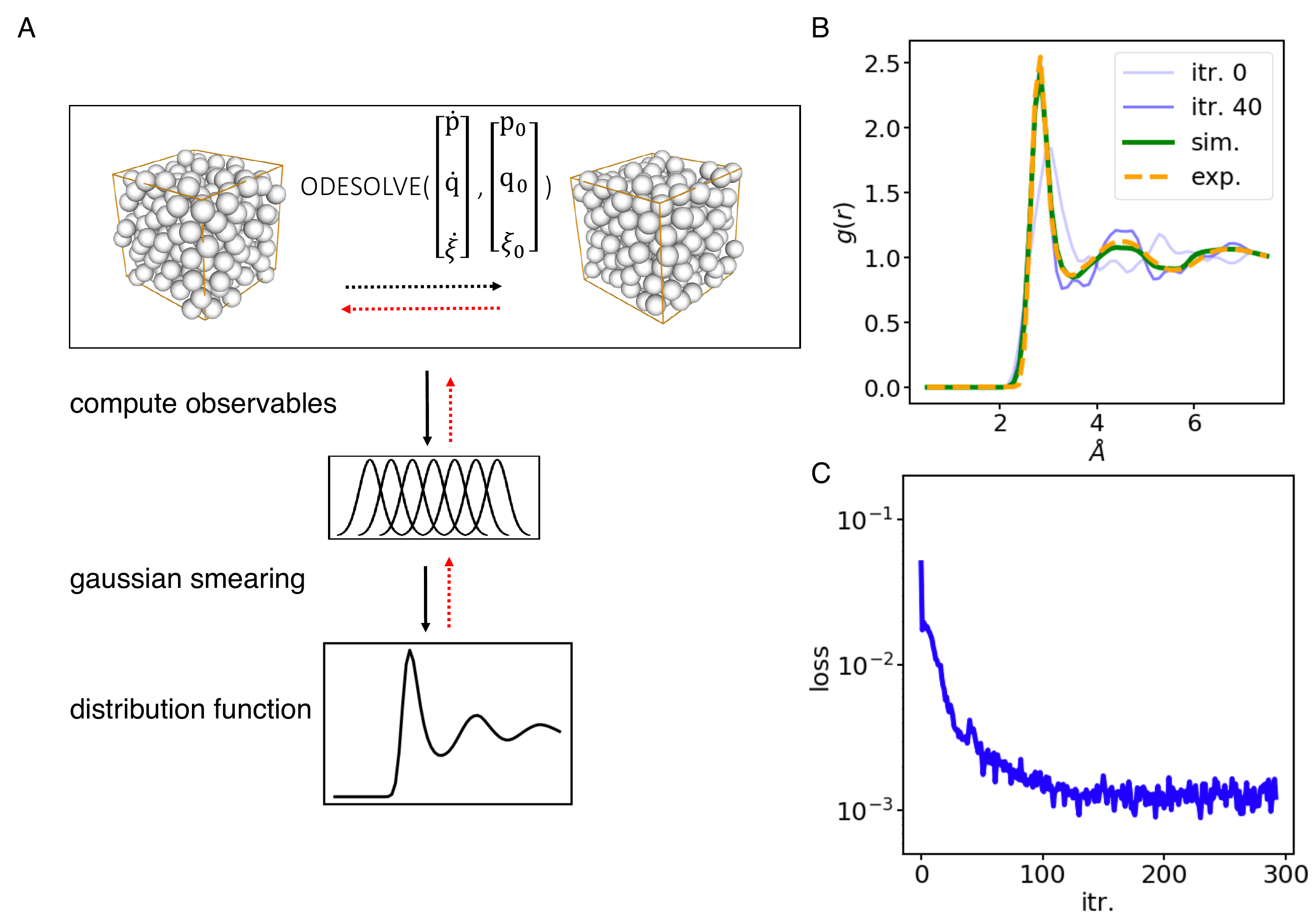}
  \caption{Computational workflow to fit pair distribution functions for. The target distributions functions are obtained from Ref.[\cite{reddy2016accuracy}] using X-ray diffraction techniques. For each training epoch, we compute the pair distribution functions from simulated trajectories. We back-propagate the mean square loss between simulated and target pair distribution functions to update the GNN parameters.}
  \label{fig:LJ}
\end{figure}

We demonstrate an example of fitting pair correlation distributions of a liquid system (Fig. \ref{fig:LJ}). Pair correlation distributions characterize structural and thermodynamic properties of condensed phase systems. \cite{Jones1924} We demonstrate that by differentiating through the simulation trajectories, one can actively modify parameters to match a target distribution function. To make the distribution function our differentiable target, we implement a differentiable histogram to approximate the typical non-differentiable histogram operation. This is done by summing over pair distances expanded in a basis of Gaussians (``Gaussian smearing"), followed by normalization to ensure that the histogram integration yields the total number of pair distances. Given an observation of pair distance $r$ that we wish to approximately map onto a histogram with $k$ bins, the Gaussian-smeared function $\rho_k(r)$ is given by

\begin{equation}
    \rho_k(r) = \frac{e^{- \frac{ (r - r_k)^2}{\delta}}}{ \sum_{k'} e^{-\frac{ (r - r_{k'})^2}{\delta}}},
\end{equation}

where $\delta$ approximates the bin width. A Gaussian basis is used to replace the non-differentiable Dirac Delta function to compute an approximate histogram. The total normalized histogram is the expected value over individual samples of $r$ over the all the observations of pair distances (between atoms $i$ and $j$) in the trajectory: 

\begin{equation}
    p(r) =  \E_{i,j} [\rho (r_{ij})].
\end{equation}

The pair correlation function $g(r)$ is obtained by normalizing by differential volumes: 

\begin{equation}
    g(r) =  \frac{V}{N^2 4 \pi r^2} p(r).
\end{equation}

We set up the experiment to learn a force field that reproduces the structure of water at 298K and 1 a.t.m which has been a challenging inverse problem. We use a Graph Neural Network as a learnable potential. the mean square loss between the simulated pair distribution function and the target. At each epoch, the parameters are updated with gradient descent. Figure \ref{fig:LJ} B shows the evolution of the simulated pair correlations during training. The loss decreases over time (\ref{fig:LJ} C) .

\section{Control protocol for light-driven quantum dynamics}

\begin{figure}[ht]
    \centering
    \includegraphics[width=4in]{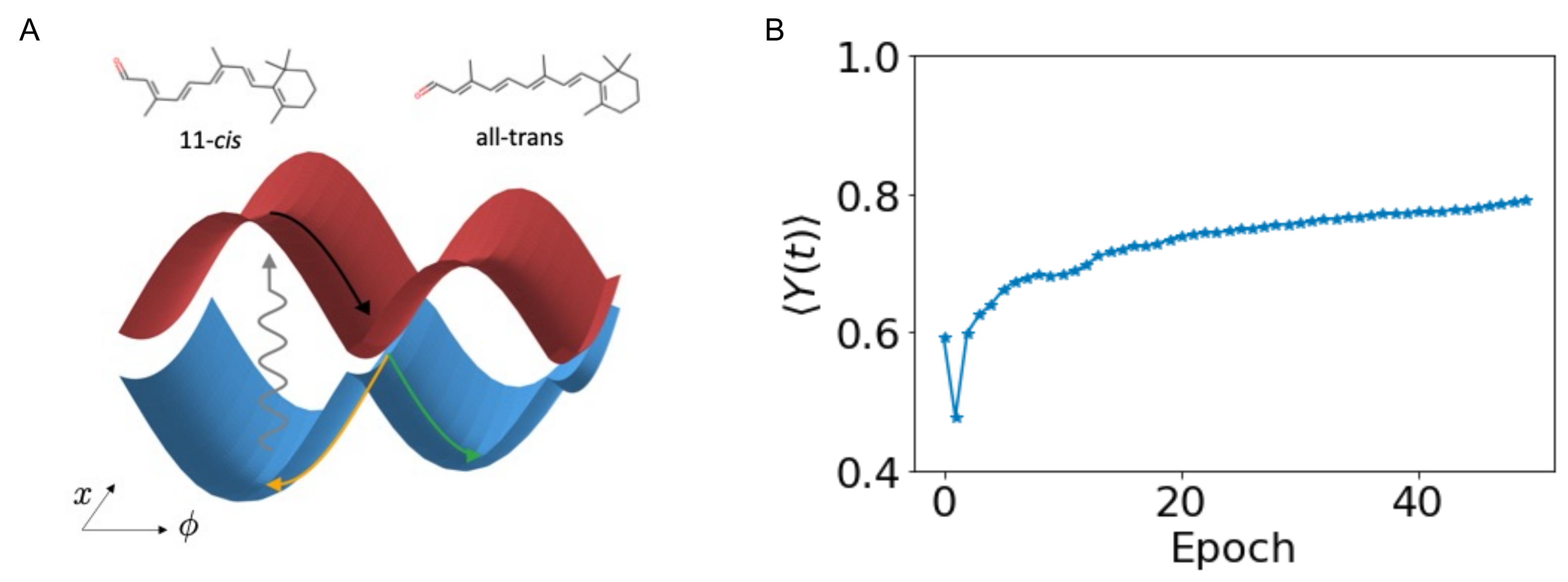}
	\caption{\textbf{A} Potential energy surfaces of the model retinal Hamiltonian. The model consists of two electronic states, denoted with blue and red, a vibrational mode $x$, and a torsional isomerization mode $\phi$. \textbf{B}  Control of the time-averaged quantum yield as a function of training epoch.}
	\label{fig:iso}
\end{figure}

We use the model introduced in Ref. \cite{hahn_stock} for the retinal chromophore. The model Hamiltonian consists of two diabatic electronic states, a single torsional mode $\phi$ for the isomerizing double bond, and a single stretching mode $x$ (see Fig. \ref{fig:fold}B). Details of the model, the construction of the Hamiltonian and the operators of interest can be found in Refs. \cite{hahn_stock, tscherbul_2, simon_2}.

The total Hamiltonian of the system and the control field is given by
\begin{align}
\hat{H}(t) = \hat{H}_{\mathrm{S}}  + \hat{H}_{{b}}(t) = \hat{H}_{\mathrm{S}} - \hat{\mu} E(t),
\end{align}
where $\hat{H}_{\mathrm{S}}$ is the system Hamiltonian, $\hat{H}_b$ is the control Hamiltonian, $\hat{\mu}$ is the dipole operator, and $E(t)$ is the electric field. The system wave function evolves under the Schr\"odinger equation,
\begin{align}
\frac{\partial}{\partial t} \ket{\psi(t)} = -\frac{i}{\hbar} \hat{H}(t) \ket{\psi(t)} \label{eq:psi},
\end{align}
where $\ket{\psi(t)}$ is the wave function and $\hbar$ is the reduced Planck constant. Computationally, the wave function is represented as a vector and the Hamiltonian as a matrix, and the system dynamics are obtained by solving Eq. (\ref{eq:psi}). Physical quantities are obtained as expectation values of operators, $A(t) = \braket{\psi(t) \vert \hat{A} \vert \psi(t)}$, where $A(t)$ is an observable and $\hat{A}$ is the corresponding operator. The minimal model contains only two nuclear degrees of freedom, but coupling to other nuclear modes and to the solvent can have significant effects on the quantum yield.\cite{balzer} These effects can be approximately treated with perturbative master equations. \cite{tscherbul_2, simon_1, simon_2} In this case a similar control scheme could be used, with the Schr\"odinger equation replaced by a master equation, and the wave function replaced by a density operator.

In the first training epoch, a Gaussian form is used for the controllable electric field:
\begin{align}
E(t) = E_0 \ \mathrm{cos}(\omega_0  (t-t_p))  \  \mathrm{exp}(-(t-t_p)^2 / \tau^2).
\end{align}
Here, $E_0$ is the amplitude of the field, $\omega_0$ is the center frequency, $t_p$ is the pulse arrival time, and $\tau$ is the pulse duration. The pulse duration is set to $\tau=10$ fs, the center frequency to $\omega_0 = 2.4$ eV, and the arrival time to $t_p = 3\tau = 30$ fs. The short pulse duration is chosen to approximate a delta-function, the arrival time to ensure that the pulse is completely contained within the simulation, and the center frequency to approximately match the electronic excitation energy. The field amplitude is initialized as $\mu E_0 /\hbar = \sqrt{\pi}/\tau$. This choice is arbitrary since, as explained below, the quantum yield is normalized with respect to excited state population, and hence to the field intensity.

The quantity to be optimized is the quantum yield, i.e. the efficiency of isomerization. The \textit{cis} projection operator is denoted as $\hat{P}_c$ and the \textit{trans} projection operator as $\hat{P}_T$. In position space, the operators are given by $\hat{P}_c = \theta(\pi/2 - \vert \phi \vert)$ and $\hat{P}_T = \theta(\vert \phi \vert -\pi/2)$, where $\theta$ is the Heaviside step function and $\phi$ the molecular rotation coordinate. The quantum yield $Y(t)$ is given by

\begin{align}
Y(t) = \frac{\braket{\hat{P}_T^{11} (t) }} { \braket{\hat{P}_T^{11}(t) } + \braket{\hat{P}_C^{00}(t) } - p_g(t) },
\end{align}
where $\braket{...} = \braket{\psi \vert ... \vert \psi}$ denotes a quantum expectation value, $\hat{P}_T^{11} = \hat{P}_T \ket{\psi_{e_1} } \bra{\psi_{e_1} }$ is the projection of $\hat{P}_T$ onto the diabatic excited electronic state, $\hat{P}_C^{00} = \hat{P}_C \ket{\psi_{e_0} } \bra{\psi_{e_0} }$ projects onto the ground diabatic state, and $p_g$ is the ground state population. Subtraction of $p_g$ ensures that any population remaining in the ground state does not contribute to the quantum yield. Since the quantum yield depends on time, we optimize its average over a time period after the pulse is over:
\begin{align}
\braket{Y(t)}_t = \frac{1}{T} \int_{t_0}^{t_0 + T}  dt \ Y(t),
\end{align}
where $\braket{...}_t$ denotes a time average. Here, $t_0$ is the time at which the yield is first recorded, and $T$ is the averaging time. We set $t_0 = 0.5$ ps and $T= 1.5$ ps. 

The dynamics are discretized with a timestep of $\Delta t = 0.05$ fs. The electric field is discretized with a time step of $5 \Delta t = 0.25$ fs. This is done to limit the timescale over which the electric field can vary. To avoid the use of complex numbers in backpropagation, the real and imaginary parts of the wave function are stored as separate vectors $\ket{\psi_R(t)}$ and $\ket{\psi_I(t)}$, respectively. They then follow the coupled equations of motion
\begin{align}
    & \frac{\partial}{\partial t} \ket{\psi_R(t)} = \frac{1}{\hbar} \hat{H} \ket{\psi_I(t)}, \nonumber \\
    & \frac{\partial}{\partial t} \ket{\psi_I(t)} = -\frac{1}{\hbar} \hat{H} \ket{\psi_R(t)}.
\end{align}
The expectation value of an operator $\hat{A}$ is given by
\begin{align}
    & \braket{\psi(t) \vert \hat{A} \vert \psi(t)} = \braket{\psi_R(t) \vert \hat{A} \vert \psi_R(t)} + \braket{\psi_I(t) \vert \hat{A} \vert \psi_I(t)} \nonumber \\
    & + i \ (\braket{\psi_R(t) \vert \hat{A} \vert \psi_I(t)} -
    \braket{\psi_I(t) \vert \hat{A} \vert \psi_R(t)}.
\end{align}
During simulations, we backpropagate through the simulation trajectory to optimize both the magnitude and phase of the temporal electric field. The numerical results are shown in Fig. \ref{fig:fold} B. The quantum yield begins at approximately 0.6, and after 50 epochs reaches 0.8 as the electric field is improved. These results show that differentiable simulations can be used to learn control protocols for electric field-driven isomerization.

\section{Conclusions}
Simulating complex landscapes of molecular states is computationally challenging. Being able to control molecular simulations allows accelerating the exploration of  configurational space and to intelligently design control protocols for various applications. In this work, we proposed a framework for controlling molecular simulations based on macroscopic quantities to develop learning and control protocols. The proposed approach is based on model learning through simulation time feedback from bulk observables. This also opens up new possibilities for designing control protocols for equilibrium and non-equilibrium simulations by incorporating bias Hamiltonians. This work can be extended to the simulation of other types of molecular systems with different thermodynamic boundary conditions and different control scenarios. 

\section{Related work}

\textbf{On differentiable simulations} Several works have incorporated physics-based simulations to control and infer movements of mechanical objects. These are done by incorporating inductive biases that obey Hamiltonian dynamics. \cite{Greydanus2019, Sanchez-Gonzalez2019} Many works also focus on performing model control over dynamical systems.  \cite{Li2019, Battaglia2016learnphysics, Zhong2019, Morton2018} Differentiable simulations/sampling with automatic differentiation have also been utilized in constructing models from data in many differential equation settings like computational fluid dynamics, \cite{Schenck2018, Morton2018} physics simulations, \cite{Hu2019chainqueen, Bar-Sinai2019, Hu2019, Liang2019cloth} quantum chemistry, \cite{Tamayo-Mendoza2018}, tensor networks \cite{Liao2019}, generating protein structures, \cite{JohnIngraham019, Ingraham2019graph, Townshend2019, Anand2019, Senior2020} and estimating densities of probability distributions with normalizing flows \cite{Zhang2018b, Grathwohl2019} and point clouds. \cite{Yang2019} Much progress has been made in developing differentiable frameworks for molecular dynamics, \cite{Schoenholz2019} PDEs, \cite{Han2018pde, Long2017, Long2019, Lu2019} and ODEs. \cite{Chen2018}

\textbf{On statistical physics and molecular dynamics} In machine learning for molecular dynamics, automatic differentiation has been applied in analyzing latent structure of molecular kinetics, \cite{Mardt2018, Xie2019, Wu2018,  Li2019flow, Wehmeyer2018, Ceriotti2019, Post2019noeq, Wang2019VAE, Hernandez2018, Chen2018md} fitting models from quantum chemistry calculations \cite{Schutt2017schnet, Bartok2010, Yao2018, Behler2007, Mailoa2019, Zhang2018, Smith2017} or experimental measurements, \cite{Xie2019foldML, Xie2018foldML} learning model reduction of atomistic simulations \cite{Lu2019, Ma2018, John2017, Wang2018Coarse, Wang2018Machine, Zhang2018CG, Bejagam2018, Durumeric2019, Hoffmann2019} and sampling microstates. \cite{Noe2018, Bonati2019, Guo2018, Rogal2019, Schneider2017} For the computation of free energy in simulations, adaptive methods and variational methods have been proposed with bias Hamiltonians on a specified reaction coordinate, \cite{Darve2008} or invertible transformations \cite{Jarzynski2001, Vaikuntanathan2008} between initial and target configurations. For non-equilibrium simulations, variational methods have been applied in the computation of large deviation functions to better sample rare events. \cite{Nguyen2016, Das2019, Dolezal2019} Studying optimal control protocols has also been a recent focus for several physical systems. \cite{Bukov2017, Rotskoff2017, Cavina2018, Cavina2018a}

\section*{Acknowledgments}
WW thanks Toyota Research Institute, SA thanks the MIT Buchsbaum Fund, and RGB thanks DARPA AMD, MIT DMSE and Toyota Faculty Chair for financial support.

\bibliographystyle{naturemag}
\bibliography{references}

\section{Appendix}

\subsection{Nose-Hover chain integrator}
Here we describe the Nose-Hover Chain \cite{Nose1984, Martyna1992}, the constant temperature integrator algorithm mentioned in the paper. We applied this integrator to the coarse-grained water and polymer examples to simulate systems with constant temperature control. Here we define the variables used in the integrator:

\begin{itemize}
\item $N$: number of particles 
\item $K$: number of virtual variables used in the chain 
\item $i$: index for individual degrees of freedom, $i: 1,  ...,  3N$
\item $j$: index for virtual variables in the chain $j: 1, ..., K$
\item $p_i$: momentum for each degree of freedom $i$
\item $q_i$: position for each degree of freedom $i$
\item $m_i$: mass for each particle in the simulation
\item $Q_j$: coupling strengths to the heat baths variable in the chain
\item $\eta_j$: virtual momenta
\end{itemize}

The coupled equations of motion are: 

\begin{equation} \label{eq:NHC}
 \begin{gathered}
 \frac{dp_i}{dt} = - \frac{\partial H}{\partial q_i}, \\
 \frac{dq_i}{dt} = \frac{\partial H}{\partial p_i} - p_i \frac{\eta_1}{Q_1} \\
 \frac{d \eta_1}{dt} = \big(\sum_i ^ {3N} \frac{p_i ^2}{2 m_i} - N k_B T \big) - \eta_1 \frac{\eta_2}{Q_2} \\
 ...\\
 \frac{d \eta_j}{dt} = \big(\frac{\eta_{j-1}}{Q_{j-1}} - k_B T \big) - \eta_j \frac{\eta_{j+1}}{Q_{j+1}} \\
 ...\\
 \frac{d \eta_K}{dt} = \big(\frac{\eta_{j-1}}{Q_{j-1}} - k_B T\big)
 \end{gathered}
\end{equation}

The Nose Hover Chain integrator performs effective temperature control, and the integrated dynamics sample the Boltzmann distribution. The integrator is deterministic and time-reversible. The control of other thermodynamic variables can be realized with other integrator protocols, such as the Rahman-Parrinelo method to maintain constant pressure. \cite{Parrinello1982}

\subsection{Learned Control Hamiltonian for Two-state Isomerization}
\begin{figure}[ht]
  \centering
  \includegraphics[width=5.5in]{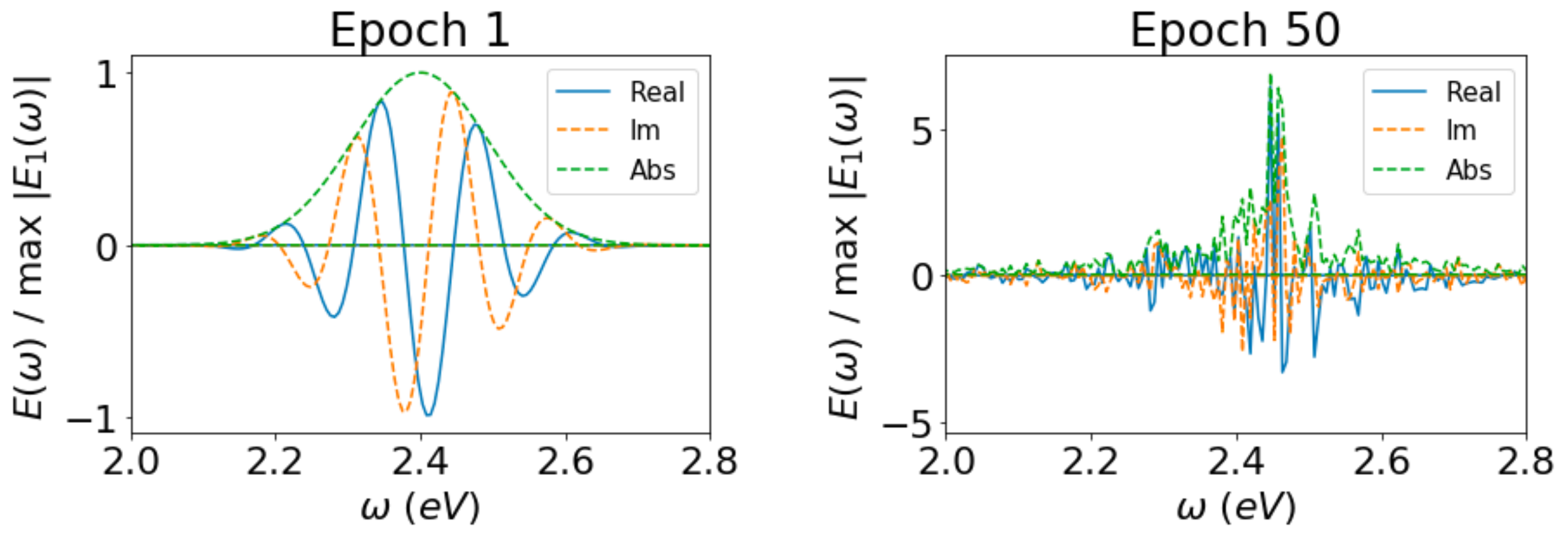}
  \caption{The comparison between the initialized electric field (epoch 1) with the learned electric field frequency (epoch 50).}
  \label{fig:e_field}
\end{figure}

We provide the learned electric field spectrum for the electric field driven two-state isomerization example in figure \ref{fig:e_field}. 

\subsection{Graph neural networks}
The model is based on graph convolution, which has achieved state-of-the-art predictive performance for chemical properties and molecular energies/forces. \cite{duvenaud_convolutional_2015, Schutt2017tensor, Zhang2018, Yao2018, Mailoa2019}. In our work, we utilized the SchNet \cite{Schutt2017schnet} architecture to learn the control Hamiltonian. The model consists of a message step and update step to systematically gather information from neighboring atoms. A 3D molecular graph is used, and the existence of a connection between atoms is decided by a fixed distance cutoff. Defining $v$ as  the index for each atom and its neighbors as $N(v)$, the graph convolutions process iteratively updates the atomic embedding $h_v$ by aggregating ``messages" from their connected atoms $v$ and their edge features $e_{uv}$. This update process is summarized by

\begin{equation} \label{eq:gcn}
	h_v^{t} = h_v^{t-1} + \sum_{u \in N(v)} Message^{t}(h_u , e_{uv}).
\end{equation}
By performing this operation several times, a many-body correlation function can be constructed to represent the potential energy surface of a molecular system. In the case of SchNet, the update function is simply a summation over the atomic embeddings. The message function is parameterized by the following equations:

\begin{equation} \label{eq:message}
 Message^{t}(e_{uv}, h_v) = MLP_3( MLP_1(e_{uv}) \circ  MLP_1(h_v)
\end{equation}

where the  $MLP_i$ are independent multi-layer perceptrons (MLP). For each convolution $t$, a separate message function is applied to characterize atom correlations at different scales. After taking element-wise products of atomic fingerprints $h_v$ and pair-interaction fingerprints $e_{uv}$, the joint fingerprint is further parameterized by another MLP to incorporate more non-linearity into the model. The final updated fingerprints are used as inputs to two fully-connected layers that yield atom-wise energies. The sum of these energies gives the total energy of the system. The atomic forces are the negative gradients of the energy with respect to atomic positions. They are easily computed through automatic differentiation implemented in PyTorch \cite{Paszke2019} and Tensorflow. \cite{Abadi2016} We used PyTorch in our demonstrations. 

\end{document}